\RequirePackage[l2tabu, orthodox]{nag}
\documentclass{stsci_report}
\usepackage[pdftex,
pdfauthor={A. M. Guzman},
pdftitle={Principal Component Analysis for ACS/WFC Superbias},
pdfsubject={},
pdfkeywords={ACS, CCD, Hubble Space Telescope, HST, Space Telescope Science Institute, STScI, Advanced Camera for Surveys, Superbias, Wide Field Channel, WFC, PCA, CTE, Principal component analysis}]{hyperref}
\usepackage[section]{placeins}
\usepackage{booktabs}
\usepackage{rotating}
\usepackage{natbib}
\usepackage{url}
\usepackage{graphicx}
\usepackage[font=footnotesize,labelfont=bf]{caption}
\usepackage{subcaption}
\usepackage{nameref}

\usepackage{array}
\usepackage{color}
\usepackage{amsmath}
\usepackage{amssymb}
\usepackage{gensymb}
\usepackage{caption}
\usepackage[bottom,symbol]{footmisc}

\usepackage{float}
\usepackage{aliascnt}
\newaliascnt{eqfloat}{equation}
\newfloat{eqfloat}{h}{eqflts}
\floatname{eqfloat}{Equation}
\usepackage{adjustbox}   
\usepackage{threeparttable}
\usepackage{booktabs}

\copyrighttext{Copyright \copyright \the\year\ The Association of Universities for Research in Astronomy, Inc. All Rights Reserved.}

\title{\textbf{Principal Component Analysis for ACS/WFC Superbias Temporal Variation}}
\presubtitle{
  \begin{flushleft}
    \includegraphics[width=5cm]{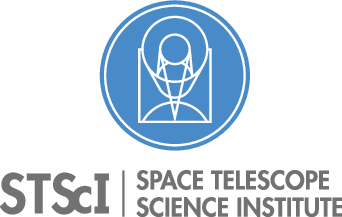}
  \end{flushleft}
  \vspace{-1.5cm}
  \begin{flushright}
    Instrument Science Report ACS 2025-03
  \end{flushright}
}

\author{A. M. Guzman, N. A. Grogin}
\date{October 21, 2025}

\begin{document}

\maketitle

\abstract{We examined the long-term behavior of the superbias calibration frames for the Advanced Camera for Surveys Wide Field Channel (ACS/WFC) aboard the Hubble Space Telescope (HST). Superbias frames are used to remove detector-level bias structure from science images and are currently generated after an anneal and delivered monthly. The primary goal of this study was to determine whether the frequency of superbias generation could be reduced without compromising calibration quality, potentially aligning with the Wide Field Camera 3 UVIS (WFC3/UVIS) approach of generating only one superbias per year. We analyzed superbias frames produced from 2007 through 2024 to investigate whether these calibration products have changed significantly over time, and whether the frequency of superbias generation and delivery could be safely reduced without
loss of calibration accuracy. In addition to visual inspections and pixel-level comparisons, we employed Principal Component Analysis (PCA) to evaluate whether any long-term, global structure exists beneath the apparent noise in these frames. Our findings show that the superbias structure has remained fairly stable post-Servicing Mission 4 (SM4), a 15-year period, and no significant or unexpected global trends or systematic shifts were detected. However, due to unstable hot columns and increasing readout dark observed in ACS/WFC data, it is likely that these calibrations still benefit from more frequent superbias updates than the annual cadence adopted for WFC3/UVIS.}

\section{Introduction} \label{s:intro}

Every month, the Hubble Space Telescope’s (HST) Advanced Camera for Surveys/Wide Field Channel (ACS/WFC) collects dark and bias frames as part of the CCD Daily Monitor Program. These raw calibration files are processed through the reference file pipeline, which generates reference files including a superbias, a non-CTE-corrected superdark, a CTE-corrected superdark, and a sink pixel file for each anneal period. These products are then submitted to the Calibration Reference Data System (CRDS), where they are used to calibrate science exposures taken during the corresponding periods.

Prior to Servicing Mission 4 (SM4), the bias structure in ACS/WFC science images was relatively uniform and stable. However, the installation of the new CCD Electronics Box Replacement (CEB-R) during SM4 introduced several features into the bias signal due to the SIDECAR ASIC \citep{sm4}. Notably, a bias gradient became apparent which caused bias levels to increase with distance from the output amplifiers in both serial and parallel directions. A low frequency $1/f$ noise component also emerged, seen as horizontal striping caused by voltage offsets applied after correlated double sampling (CDS). Additionally, in Dual-Slope Integrator (DSI) mode, a signal-dependent bias shift was observed: bias levels vary based on pixel intensity and position due to asymmetric feedback in the CDS ramping \citep{sm4}. These features are stable but require specific calibration strategies. Despite these changes, the updated electronics provided improved read noise performance in the WFC data.

Superbias frames specifically address fixed-pattern noise and baseline offsets introduced during CCD readout. As such, they are critical to the accurate reduction of ACS/WFC science data. Historically, these frames have been delivered on a monthly cadence under the assumption that bias structure may change frequently due to effects of the electronics or radiation effects. However, the Wide Field Camera 3 (WFC3) team has shown that the UVIS bias behavior remains stable enough to justify generating only one superbias per year, greatly reducing the number of internal orbits used \citep{wfc3_isr}. In the CALACS pipeline, superbias subtraction is applied in the \textbf{doBias} step, where the superbias is subtracted from raw science images before they are converted from Data Numbers (DNs) to electrons. The superbias corrects only the static bias structure and not time-dependent or readout-level variations (e.g., bias drifts, or striping). These latter effects are addressed in the \textbf{doBlev} step. Each quadrant of the ACS/WFC detector has its own readout amplifier and corresponding overscan region. The physical pre-scan columns are used to estimate the quadrant-specific bias level. However, even after superbias and bias level subtraction, small random amplifier-dependent bias offsets (typically $\sim$ 0.3 DN) may remain. These residuals can cause visible discontinuities in background levels across quadrant boundaries, particularly in low-background exposures \citep{acs_dhb}.

This project was initiated to determine whether ACS/WFC might exhibit a similar degree of stability and whether the frequency of superbias generation and delivery could be safely reduced without loss of calibration accuracy. In Section \ref{s:analysis}, we describe the data and outline our analysis strategy. The results of this analysis are presented in Section \ref{s:results}, followed by our conclusions in Section \ref{s:conclusion}.

\section{Data \& Analysis} \label{s:analysis}

\subsection{Linear Regression Analysis}
\begin{figure}[H]
	\centering
	\begin{subfigure}{1\textwidth}
		\centering
		\includegraphics[scale=0.10]{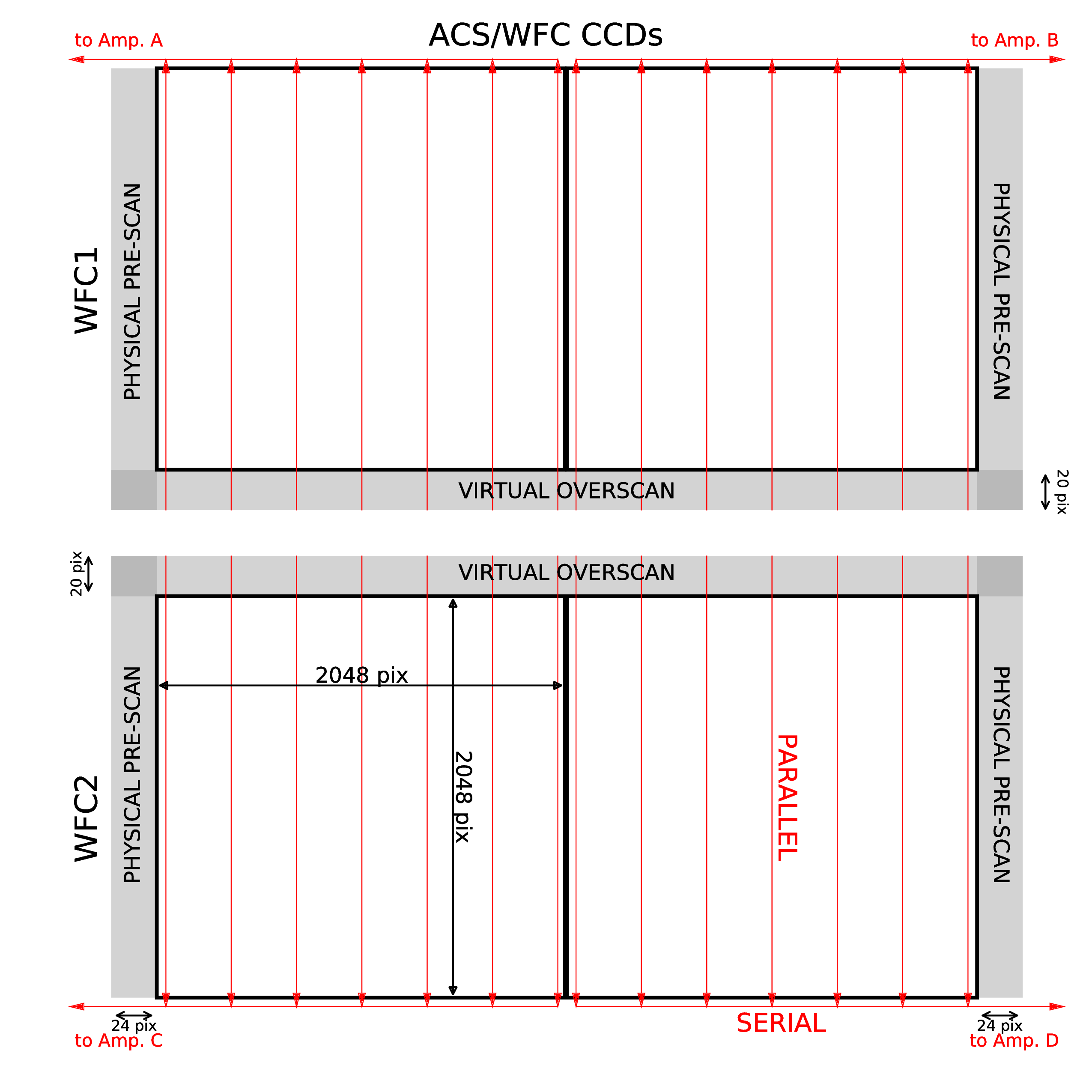}
		\end{subfigure}
\caption{The layout of ACS/WFC full-frame images, including parallel and serial readout directions, physical pre-scans, and virtual overscans \citep{acs_ihb}.}
    \label{fig:ccd}
\end{figure}

For this portion of the analysis, a total of 197 ACS/WFC superbias reference files, ranging from 2009 to 2024, were used. The analysis was confined to the science region (4096$\times$2048 pixels) for both WFC1 and WFC2 of each reference file by excluding both the physical pre-scans and virtual overscans, as seen in Figure \ref{fig:ccd}. These superbiases are stacks of $\sim$ 50 individual bias frames taken throughout the 1-month period between anneals. The bias level is removed and cosmic rays are rejected, leaving hot columns and 2D bias structure.

\begin{figure}[H]
	\centering
	\begin{subfigure}{1\textwidth}
		\centering
		\includegraphics[scale=0.32]{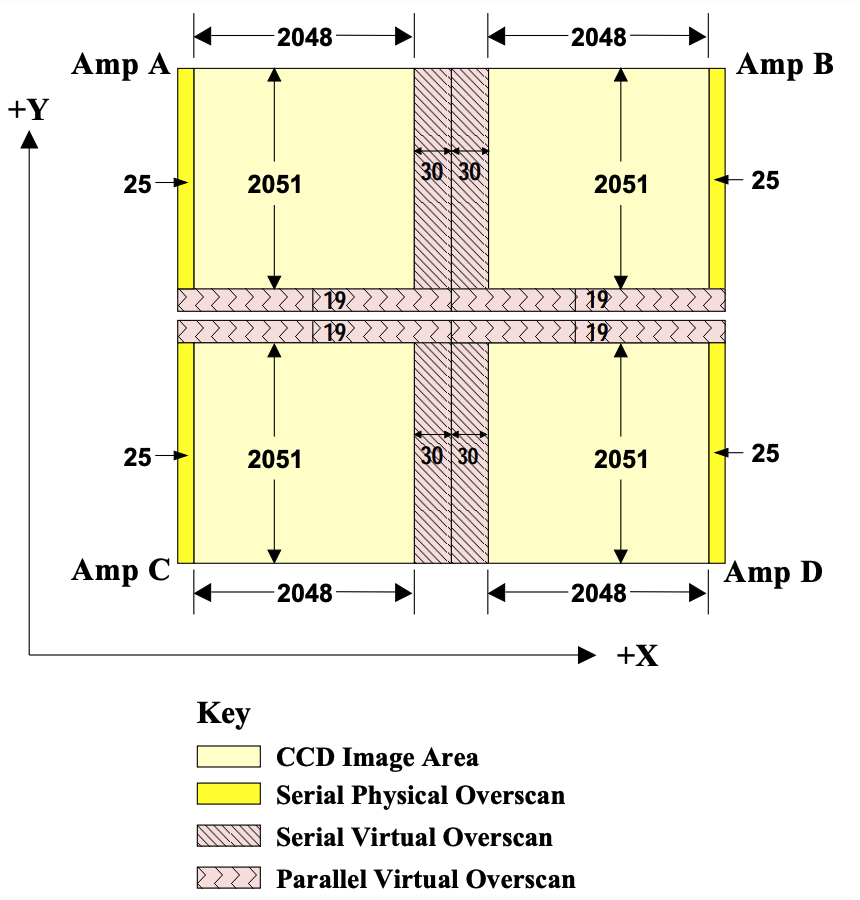}
		\end{subfigure}
\caption{The layout of the WFC3/UVIS full frame detector \citep{wfc3_ihb}.}
    \label{fig:wfc3_uvis}
\end{figure}

For comparison, 15 superbias reference files from the WFC3/UVIS detector (spanning from 2009-2022) were also analyzed. These frames were similarly trimmed to include only the science region (2051$\times$2048 pixels), excluding the serial physical overscans, serial virtual overscans, and parallel virtual overscans, as seen in Figure \ref{fig:wfc3_uvis}.

\subsection{Principal Component Analysis}

Principal Component Analysis (PCA) is a statistical technique used to reduce the dimensionality of data by transforming it into a new set of orthogonal variables, called principal components (PC), which capture the directions of maximum variance in the dataset. Each principal component corresponds to an eigenvector of the data’s covariance matrix, and the amount of variance it explains is given by the associated eigenvalue. In the context of analyzing the ACS/WFC superbias frames, PCA can be especially useful for identifying underlying patterns or trends over time. By applying PCA to a time series of superbias frames, one can determine whether significant structural or pixel-level changes are occurring in the bias pattern, such as readout dark \citep{dancoe_isr}, drifts, evolving artifacts, or the appearance of new systematic features. 

For the PCA, we included the 2007 superbias reference files to explore whether the analysis could identify differences between pre- and post-SM4. For comparison, we also simulated evolving bias structure over time, by applying a synthetic gradient to each superbias SCI array. The purpose of this simulation was to create a dataset with a smoothly evolving bias structure over time, allowing us to observe how the principal components behave under controlled, gradual changes. This gradient increases in strength with the file index, where the index corresponds to the order of frames sorted by observation date. Thus, the earliest file (from 2007) has the weakest gradient, while the final file (from 2024) has the strongest. The gradient was applied in the $x$-direction to WFC1, and $y$-direction to WFC2, where the $x$-direction gradient is proportional to $(x - 2048)/2048$ and the $y$-direction gradient is proportional to $(y - 1024)/1024$. Both gradients are scaled by a time-dependent factor defined as $(0.5 + n/N)$, where $n$ is the index of the current file and $N$ is the total number of files (see Equations 1 and 2). This allowed for a smooth increase in gradient strength across the dataset. The gradients in the $x$ and $y$ direction, $factor_x$ and $factor_y$, including both position and time dependence, are given by:

\begin{equation}
\text{factor}_x = \left(0.5 + \frac{n}{N}\right) \cdot \frac{x - 2048}{2048}
\end{equation}

\begin{equation}
\text{factor}_y = \left(0.5 + \frac{n}{N}\right) \cdot \frac{y - 1024}{1024}
\end{equation}

\begin{figure}[H]
	\centering
	\begin{subfigure}{1\textwidth}
		\centering
		\includegraphics[scale=0.22]{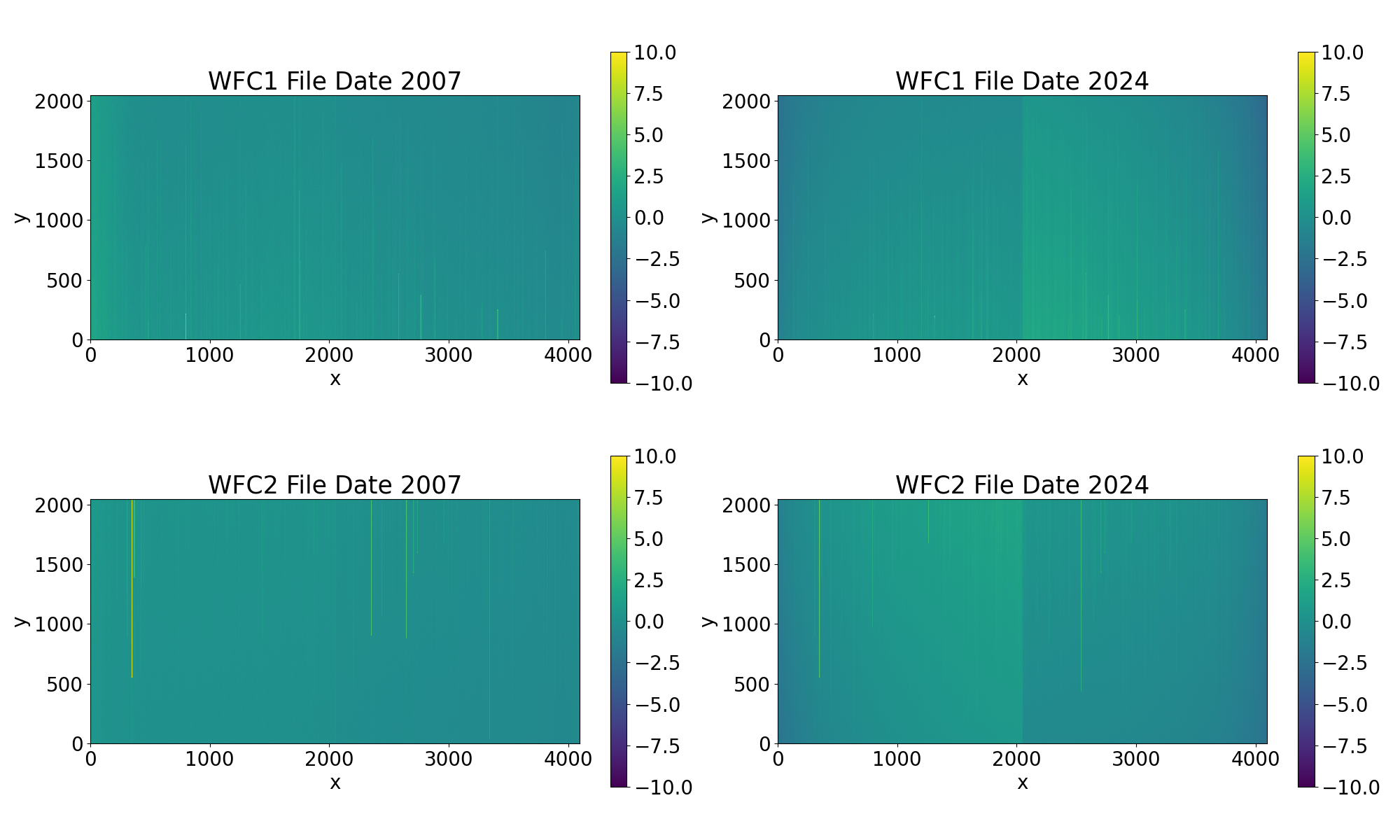}
		\end{subfigure}
\caption{First (2007) and last (2024) original standardized superbias frames used in this analysis, with WFC1 shown at the top and WFC2 at the bottom.}
    \label{fig:og_2007}
\end{figure}

\begin{figure}[H]
	\centering
	\begin{subfigure}{1\textwidth}
		\centering
		\includegraphics[scale=0.22]{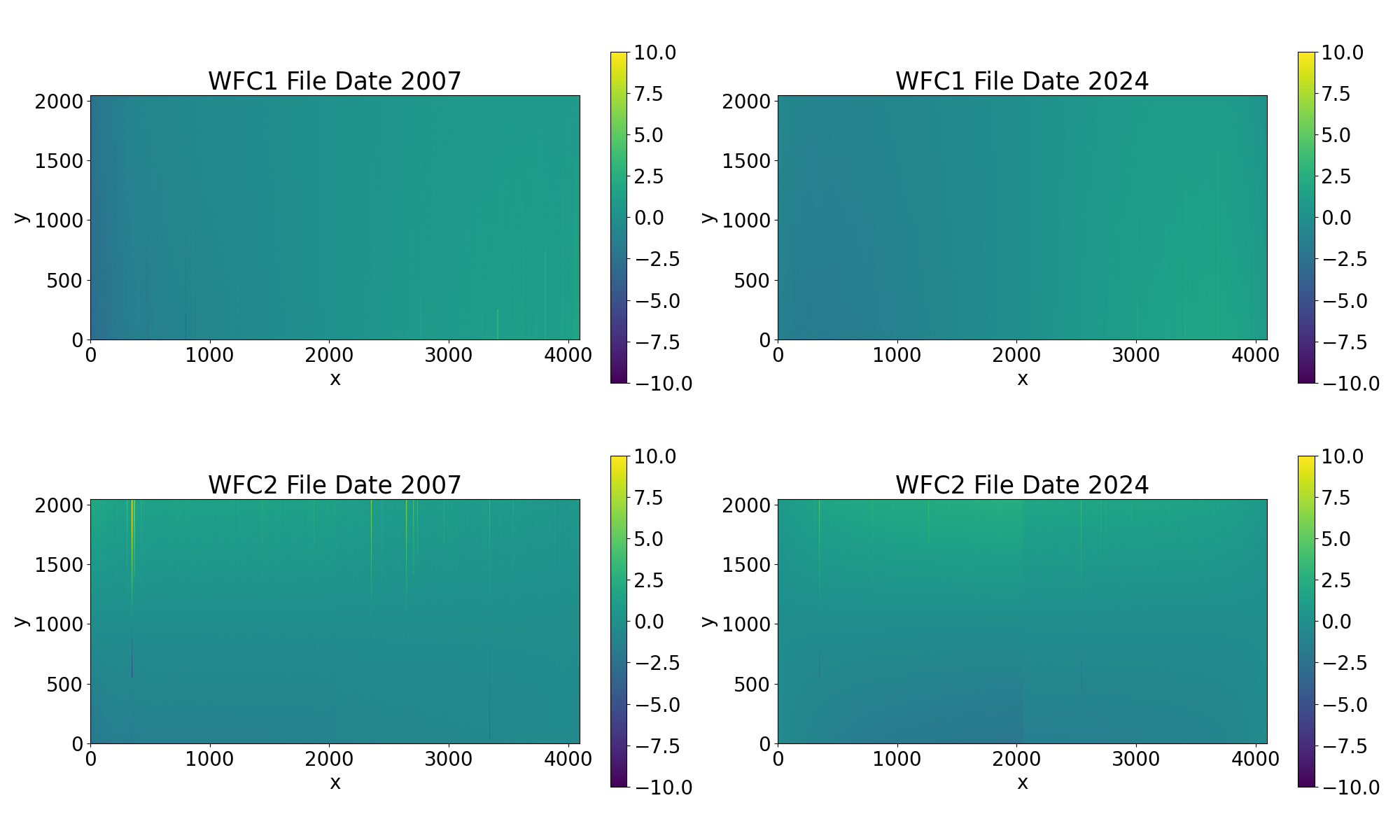}
		\end{subfigure}
\caption{First (2007) and last (2024) standardized superbias frames with gradients applied along the $x$-axis (top) and $y$-axis (bottom).}
    \label{fig:sim_2007}
\end{figure}

PCA is sensitive to differences in scale between input variables, therefore the data must be standardized before performing PCA. Each superbias frame, with and without the gradient added, was standardized by subtracting the mean and dividing by the standard deviation of the SCI array of the appropriate superbias file. This ensures that all pixels contribute equally to the analysis, regardless of their original values. Without this step, the PCA would be dominated by the highest-value pixels, such as hot pixels, potentially obscuring important but smaller-scale patterns. Figure \ref{fig:og_2007} shows the first and last standardized superbias files used in the analysis, while Figure \ref{fig:sim_2007} shows the first and last superbias with the synthetic gradient added.

To quantify how many components are needed to describe the majority of variance in the superbias frames, we used the PCA tool from the \texttt{sklearn.decomposition} package, using the criterion \texttt{n\_components=0.90}, to select the minimum number of principal components required to explain at least 90\% of the total variance. This was applied to the original superbias frames (WFC1 and WFC2) after standardization. The resulting component counts were 127 for WFC1 and 126 for WFC2, with cumulative variance explanations of 90.10\% and 90.19\%, respectively.

\begin{figure}[H]
	\centering
	\begin{subfigure}{1\textwidth}
		\centering
		\includegraphics[scale=0.11]{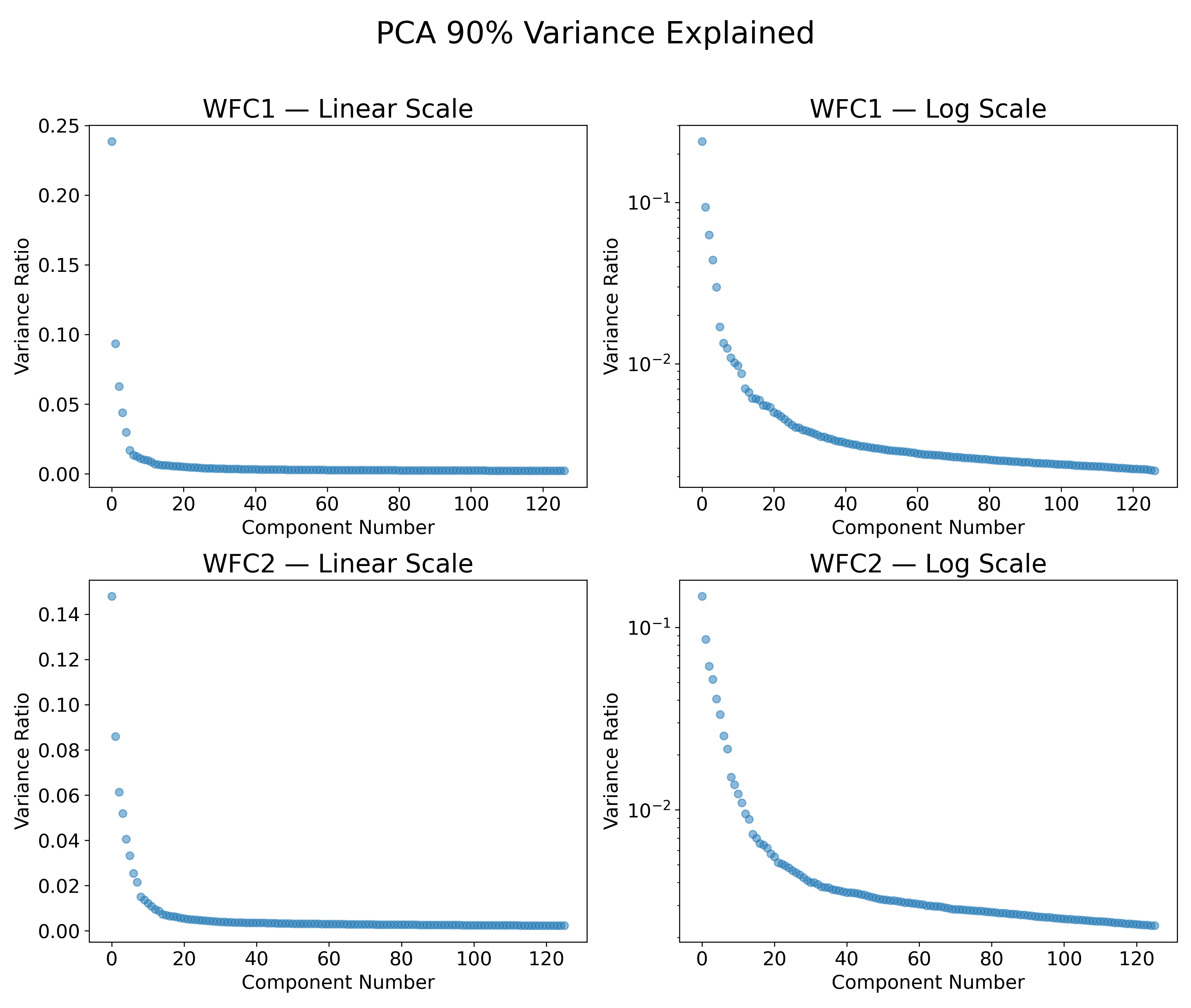}
		\end{subfigure}
\caption{PCA variance plots for the original superbias frames. Left: Variance ratio explained by each component, showing a steep drop after the first few components for both chips, followed by a flattening trend (“elbow”). Right: Same variance data on a logarithmic scale, revealing the long tail of minor variance contributions beyond the top ~20 components. There are 127 total PCs for WFC1 and 126 components for WFC2.}
    \label{fig:wfc1_wfc2_variance_og_2007}
\end{figure}

\begin{figure}[H]
	\centering
	\begin{subfigure}{1\textwidth}
		\centering
		\includegraphics[scale=0.11]{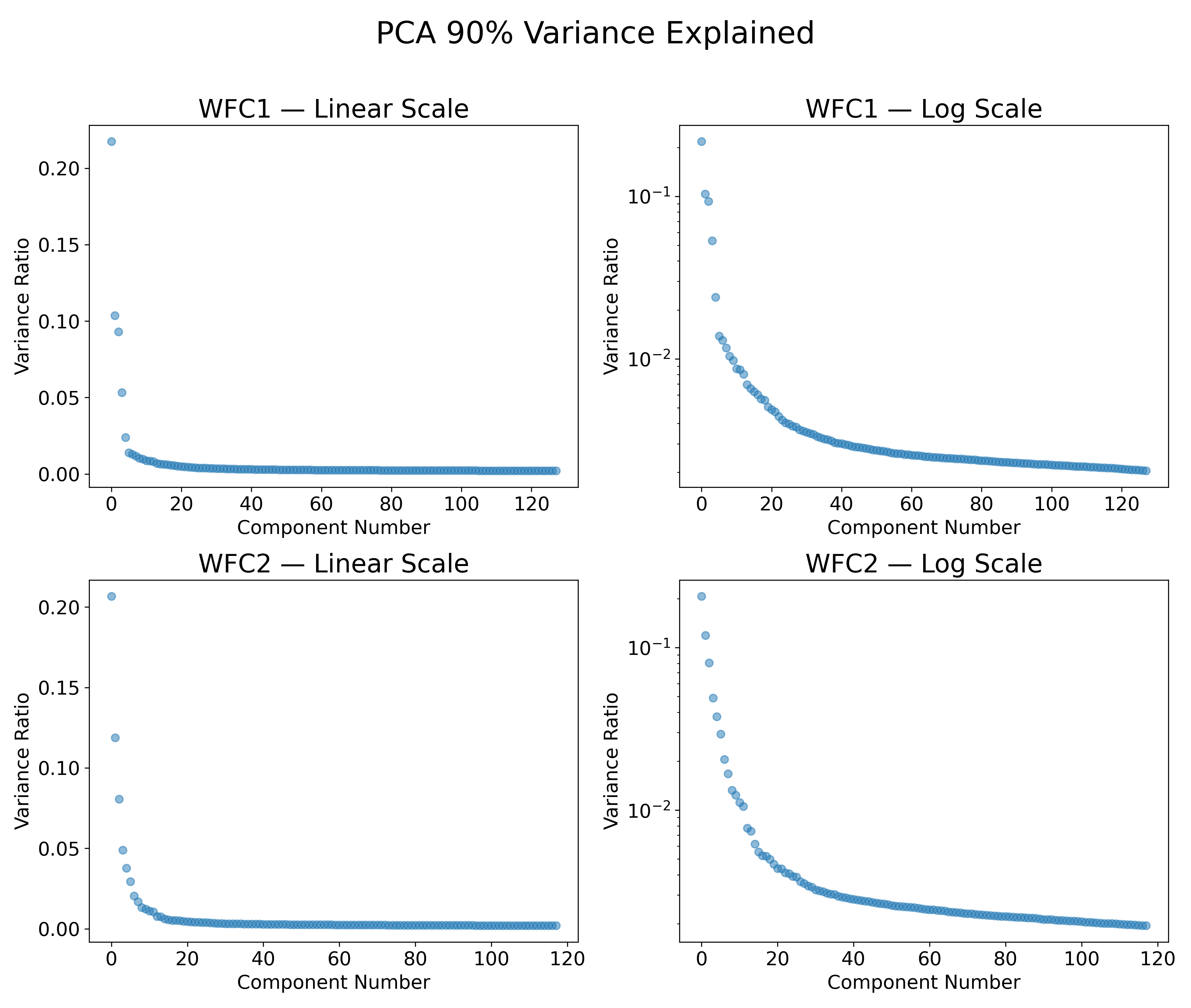}
		\end{subfigure}
\caption{Same as Figure \ref{fig:wfc1_wfc2_variance_og_2007} but for the simulated superbias frames. There are 128 total PCs for WFC1 and 118 components for WFC2.}
    \label{fig:wfc1_wfc2_variance_sim_2007}
\end{figure}

The same method was applied to the superbias frames with additional synthetic gradient, in order to calculate the number of components that explain at least 90\% of the variance. The results indicate that for WFC1, 128 principal components were required to explain 90.12\% of the total variance, while for WFC2, only 118 components were needed to reach 90.00\%. 
We then plot these results for the original superbias (Figure \ref{fig:wfc1_wfc2_variance_sim_2007}) and the superbias frames with synthetic gradient (Figure \ref{fig:wfc1_wfc2_variance_og_2007}). The linear-scale variance plot shows that the first few principal components capture only a modest portion of the total variance. While PC1 explains the largest share, subsequent components rapidly taper off in contribution. To better visualize the long tail of smaller components, we also plotted the variance ratios on a logarithmic scale. In a PCA variance plot, the ``knee" or ``elbow" refers to the point on the curve where the explained variance begins to level off. This indicates that additional components beyond this point contribute progressively less to explaining the overall variance in the dataset. For this dataset, the variance log-scale plots reveal that even beyond the first 20–30 components, a substantial number of PCs carry small meaningful amounts of variance. 

\section{Results} \label{s:results}

\subsection{Linear Regression Analysis}

\begin{figure}[H]
	\centering
	\begin{subfigure}{1\textwidth}
		\centering
		\includegraphics[scale=0.17]{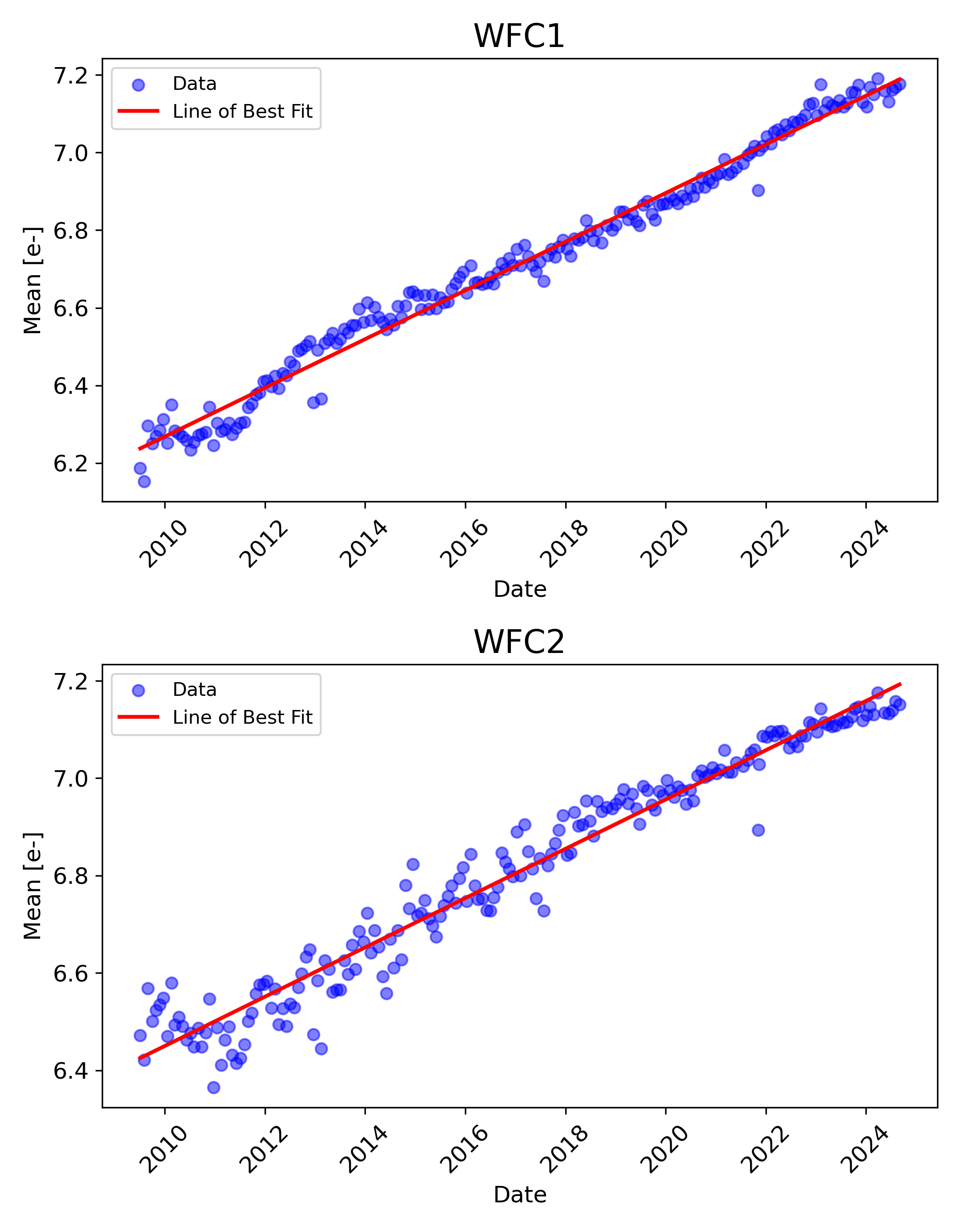}
		\end{subfigure}
\caption{The mean superbias level for ACS/WFC1 (top) and WFC2 (bottom) detectors from 2009 to 2024. Blue points represent individual superbias frames, the red line shows the best-fit linear regression. The slope for WFC1 is $\sim 0.0627$ $e^-$ per year  and for WFC2 is $\sim 0.0507$ $e^-$ per year, indicating a slow, consistent increase in bias level over time.}
    \label{fig:slope_WFC}
\end{figure}

\begin{figure}[H]
	\centering
	\begin{subfigure}{1\textwidth}
		\centering
		\includegraphics[scale=0.17]{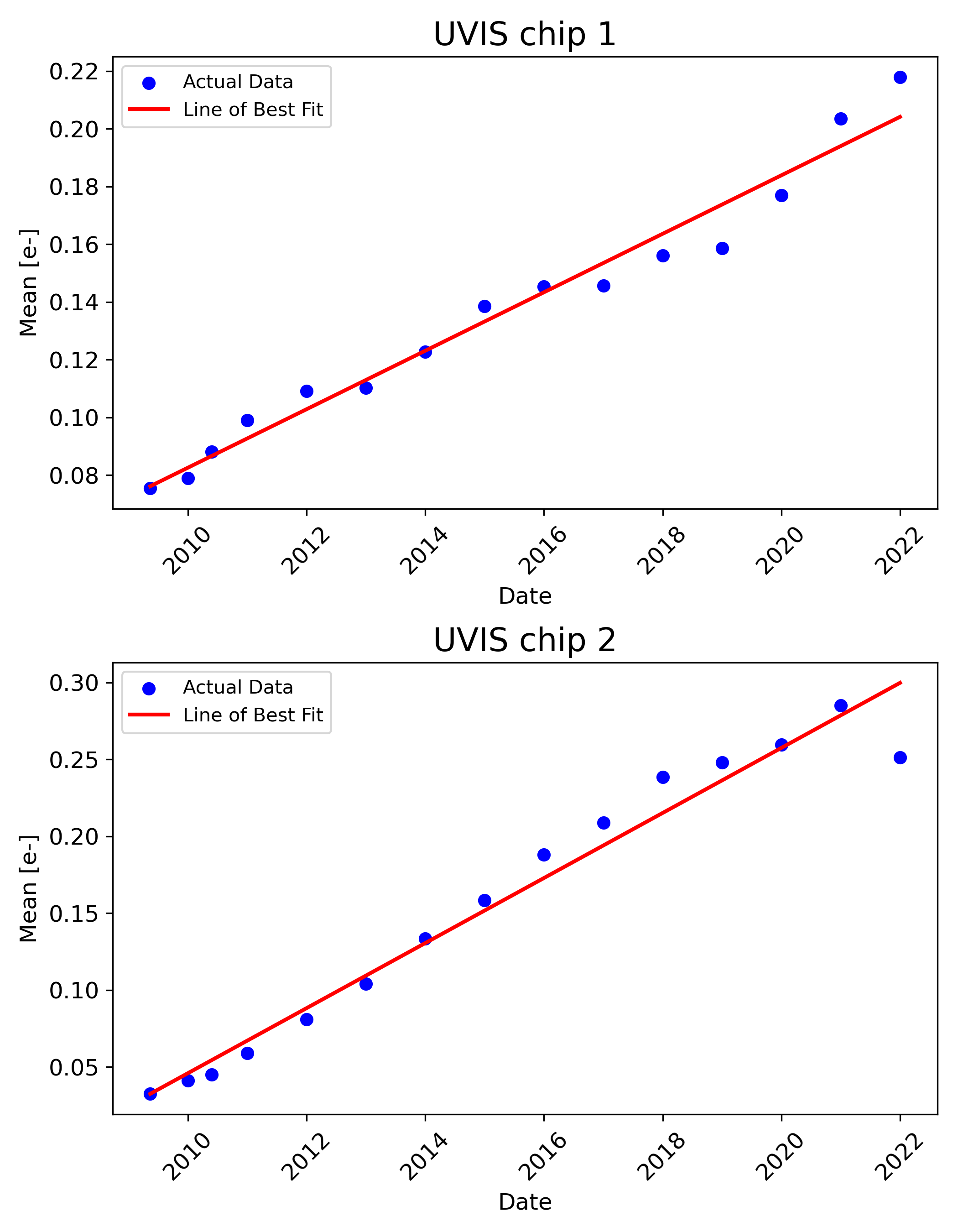}
		\end{subfigure}
\caption{The mean superbias level for WFC3/UVIS chip 1 (top) and chip 2 (bottom) from 2009 to 2022. Each blue point represents a superbias frame; the red line shows a linear fit. The slopes indicate a gradual increase of $\sim 0.010$ $e^-$ per year for UVIS1 and $\sim 0.021$ $e^-$ per year for UVIS2, pointing to long-term evolution in the detector’s electronic bias structure.}
    \label{fig:slope_uvis}
\end{figure}

The mean bias level of each chip in every superbias file was calculated and plotted in Figure \ref{fig:slope_WFC}. The mean pixel value of each cropped superbias was calculated using \texttt{numpy}, and a linear trend over time was determined using \texttt{sklearn.linear\_model.LinearRegression} to compute a line of best fit. Both WFC1 and WFC2 exhibit a significantly higher baseline and a more pronounced evolution compared to WFC3/UVIS (Figure \ref{fig:slope_uvis}). In the ACS/WFC data, the mean bias levels start around 6.2–6.4 electrons in 2009 and rise steadily to nearly 7.2 electrons by 2024. This increase is well-captured by a linear trend, with fitted slopes of $\sim 0.0627$ $e^{-}$ per year for WFC1 and $\sim 0.0507$ $e^-$ per year for WFC2. The long-term rise in bias level indicates a gradual evolution in the detector’s baseline readout level, possibly due to increasing readout dark—dark current accrued during readout. This same increase in average SCI arrays over time is also likely driven by the rise in readout dark.

In contrast, the WFC3/UVIS superbias levels are substantially lower in magnitude and show a slower increase with time. UVIS chip 1 maintains a mean bias level between 0.07 and 0.22 electrons, while chip 2 ranges from 0.03 to 0.30 electrons over the same period. The corresponding slopes are $\sim 0.010$ $e^-$ per year for UVIS1 and $\sim 0.021$ $e^-$ per year for UVIS2 roughly 3 to 6 times smaller than those of the ACS/WFC detectors. Additionally, the UVIS superbias trends in Figure \ref{fig:slope_uvis} appear smoother and more stable, with less scatter around the line-of-best-fit. This suggests that WFC3/UVIS experiences slower temporal changes that are due to dark current accumulated during readout (96 seconds) and charge transfer efficiency losses in the detector \citep{wfc3_isr2}.

Overall, this comparison highlights that while both instruments show a gradual increase in superbias level over time, the effect is more severe and pronounced in ACS/WFC. WFC3/UVIS superbias levels are not only lower, but also more stable over multi-year timescales, indicating better long-term electronic stability in the WFC3/UVIS detectors. These differences are important when considering calibration strategies, as the higher and more variable superbias structure in ACS may require more frequent updates or corrections to maintain calibration accuracy.

\subsection{Principal Component Analysis}

\begin{figure}[H]
	\centering
	\begin{subfigure}{1\textwidth}
		\centering
		\includegraphics[scale=0.32]{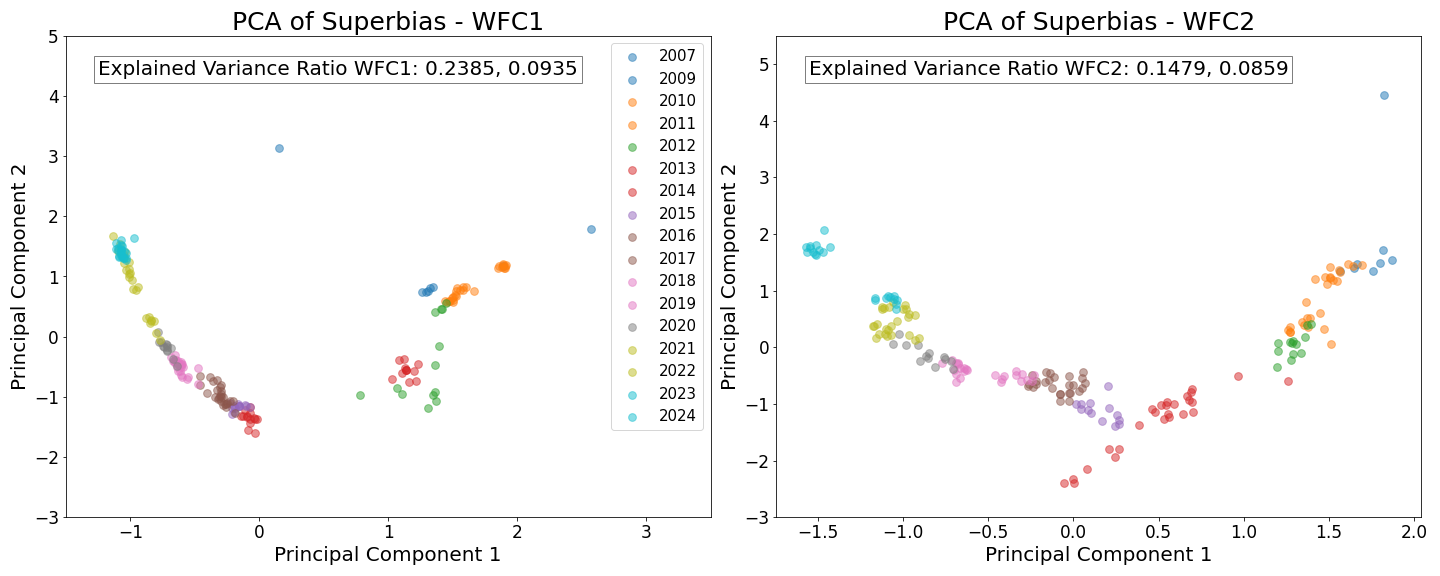}
		\end{subfigure}
\caption{PCA of ACS/WFC superbias frames from 2007 to 2024 without any artificial gradient applied. The first two principal components explain over 32\% of the total variance in WFC1 and 23.8\% of the total variance for WFC2, with PC1 alone capturing most of the structure. The 2007 frames appear as early outliers, with a PC1 and PC2 of $(0.3, 3.4)$ for WFC1 and $(1.8, 4.7)$ for WFC2, suggesting differences in detector bias structure over time.}
    \label{fig:og_pca_2007}
\end{figure}

We reshaped the first and 20th components, for both the original and simulated PCA, back to the superbias' original size (4096$\times$2048 pixels) to visualize what each component highlights in the superbias frames. The PCA results for the original superbias frames, those without any added gradient, are shown in Figure \ref{fig:og_pca_2007} color-coded by year. For WFC1 (left plot), the first two principal components (PC1 and PC2) explain 23.85\% and 9.35\% of the variance, respectively. For WFC2 (right plot), the corresponding values are 14.79\% and 8.59\%. These values suggest that a substantial portion of the total variation is captured within the first two PCs, particularly in WFC1. The data points plotted reflect moderate but orderly variation between frames, with certain years (e.g., 2009 and later) forming distinct clusters, implying that inter-annual changes in the superbias are present.  Notably, in both WFC1 and WFC2, the 2007 superbias appears as a strong outlier. Its separation in principal component space supports prior knowledge of instrumental and calibration differences before and after SM4. The strong temporal progression in PCA space confirms a global, time-dependent trend in the superbias frames.

\begin{figure}[H]
	\centering
	\begin{subfigure}{1\textwidth}
		\centering
		\includegraphics[scale=0.32]{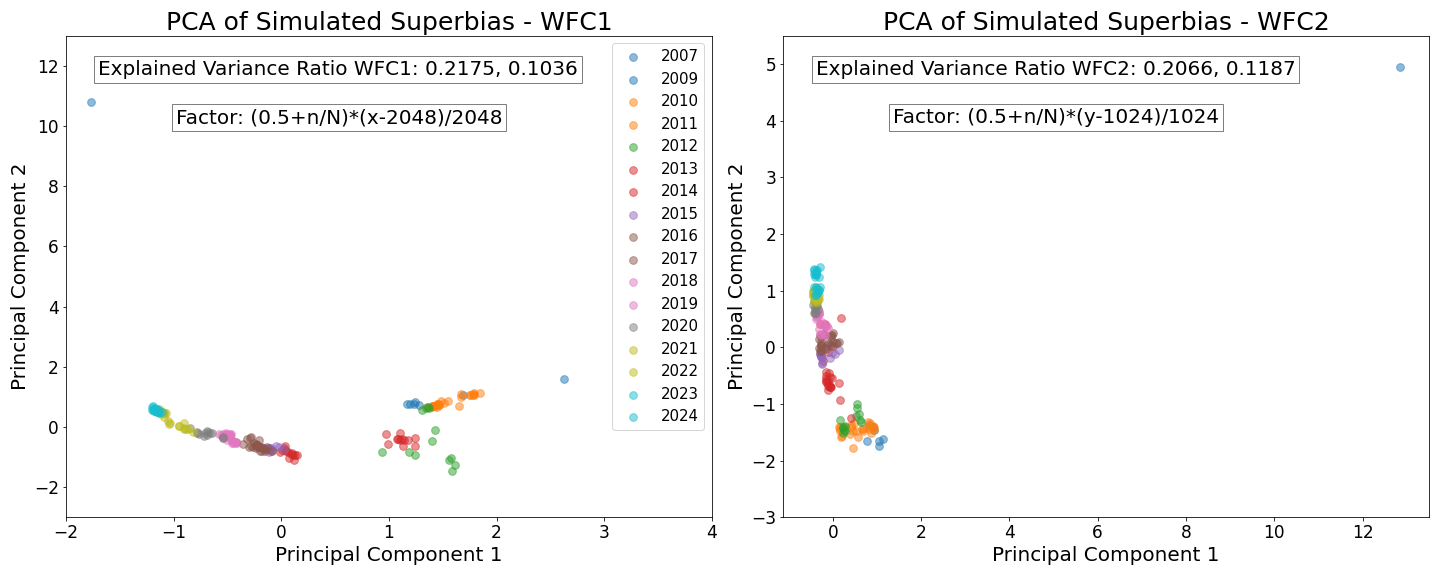}
		\end{subfigure}
\caption{PCA of ACS/WFC superbias frames from 2007 to 2024 after adding a synthetic spatial gradient: in the $x$-direction for WFC1 and $y$-direction for WFC2. The first principal component explains over 20\% of the variance in both detectors showing PCA’s sensitivity to global spatial structure. The 2007 superbias file appears as an outlier, with a PC1 and PC2 of $(-1.8, 12)$ for WFC1 and $(-13.8, 5.1)$ for WFC2.}
    \label{fig:sim_pca_2007}
\end{figure}

In contrast, the simulated data shows clear structural differences in PC space (Figure \ref{fig:sim_pca_2007}). The results reveal that for WFC1 (gradient in the $x$-direction), PC1 explains 21.75\% of the variance, while PC2 explains 10.36\%. Similarly, for WFC2 (gradient in the $y$-direction), PC1 accounts for 20.66\% and PC2 for 11.87\% of the variance. This increase in explained variance for WFC2 indicates that the applied gradient in the $y$-direction is faintly captured by the PCA. However, the same amount of total variance ($\sim$32\%) is captured in both the original and simulated analysis for WFC1. Like the original superbias PCA, in both WFC1 and WFC2, the 2007 superbias appears to be an outlier. The continued separation of the 2007 superbias frame in the PCA results, even after a gradient is applied to the data, suggests that the structural changes induced by the SM4 electronics cannot be reproduced by gradual temporal variation alone. 

\begin{figure}[H]
	\centering
	\begin{subfigure}{1\textwidth}
		\centering
		\includegraphics[scale=0.32]{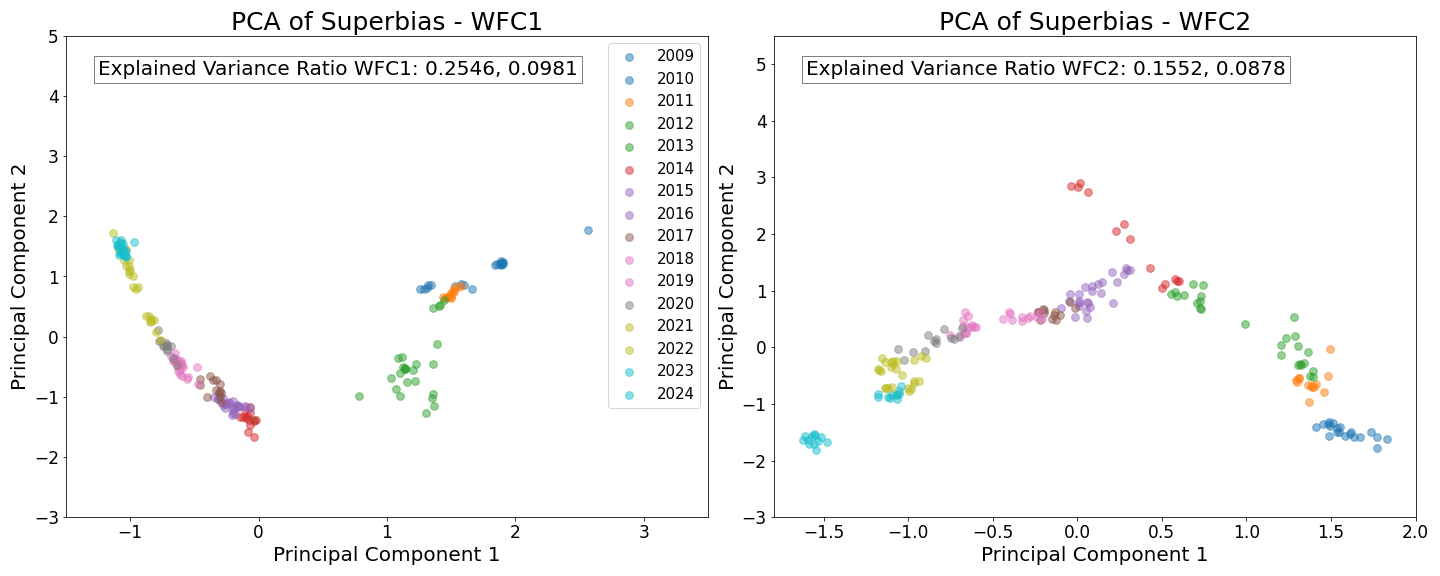}
		\end{subfigure}
\caption{PCA of the original ACS/WFC superbias frames from 2009–2024 without any added gradients. The first two principal components explain 25.46\% and 9.81\% of the variance for WFC1 (left), and 15.52\% and 8.78\% for WFC2 (right).}
    \label{fig:og_pca_2009}
\end{figure}

To further investigate the influence of the 2007 superbias outlier, a new PCA was performed using only post-SM4 ACS/WFC data, from 2009 through 2024. The resulting PCA scatter plots for WFC1 (left) and WFC2 (right) are shown in Figure \ref{fig:og_pca_2009}. The explained variance ratios of the first two components are (0.2546, 0.0981) for WFC1 and (0.1552, 0.0878) for WFC2, indicating that these two components account for approximately 35.27\% and 24.31\% of the total variance in each chip, respectively.

The results demonstrate that the temporal structure present in the superbias data remains even without the inclusion of the 2007 outlier. In WFC1, the frames exhibit a smooth progression along PC1 and PC2, with clear temporal clustering that suggests a gradual evolution in bias structure over time after 2013. A similar trend is observed in WFC2, although the distribution appears more dispersed. These patterns imply that despite the removal of the 2007 outlier, the underlying bias structure continues to change in a coherent, temporally dependent way. 

\begin{figure}[H]
	\centering
	\begin{subfigure}{1\textwidth}
		\centering
		\includegraphics[scale=0.075]{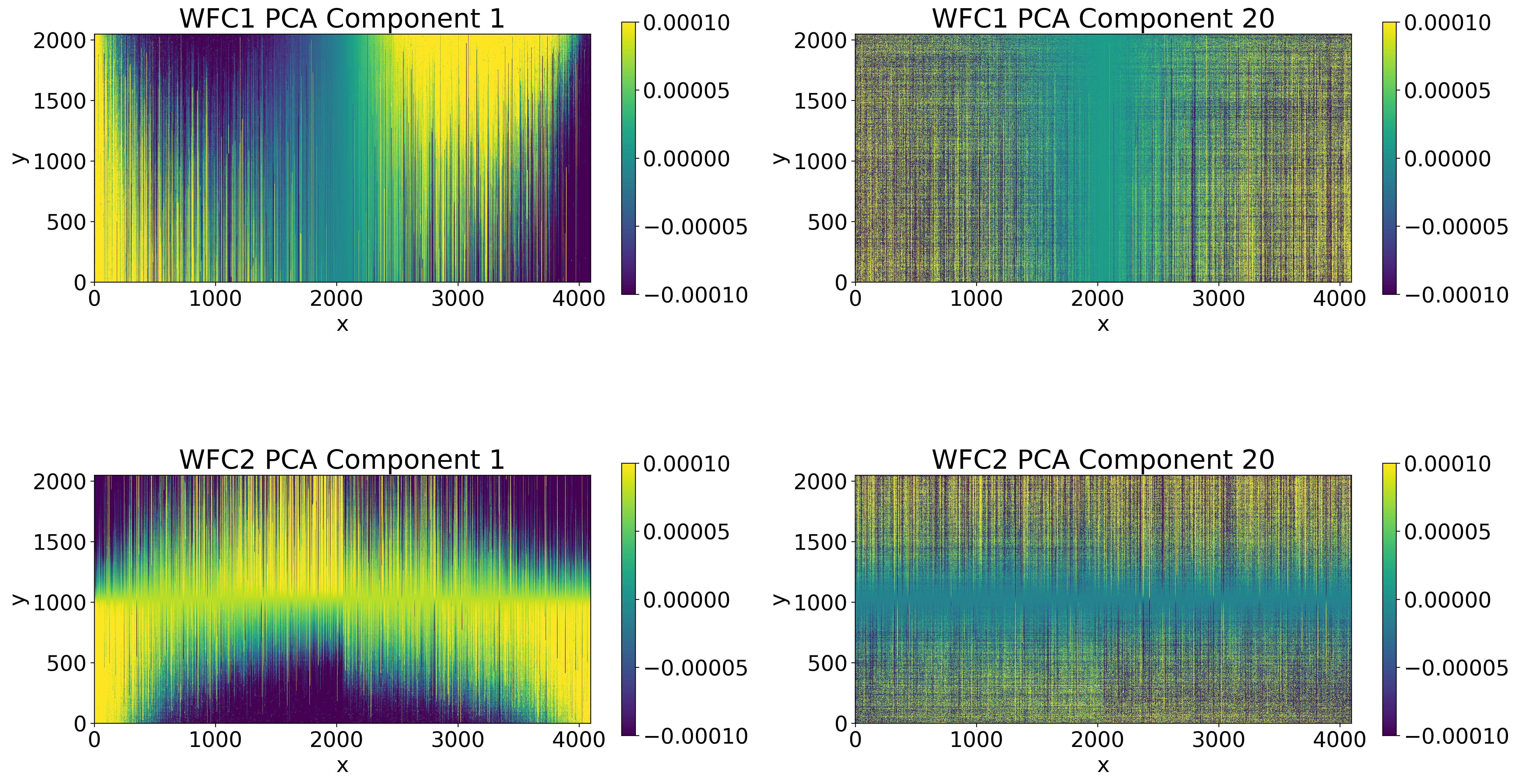}
		\end{subfigure}
\caption{Reconstructed superbias frames from selected principal components  for the simulated 2009 data with added gradients. The left column shows the first principal component, and the right column shows the 20th principal component. WFC1 reconstructions are shown in the top row and WFC2 in the bottom row. The added gradient is clearly visible in PC1 for both WFC1 and WFC2, while PC20 contains smaller changes across the detector.}
    \label{fig:wfc1_wfc2_components_sim_2007}
\end{figure}

Finally, Figure \ref{fig:wfc1_wfc2_components_sim_2007} shows the reconstructed images of the first and twentieth principal components of the WFC1 and WFC2 simulated superbias frames. The images clearly show how PCA has identified the dominant patterns in the added gradient along with hot columns in the superbias frames. For both detectors, the first principal component (PC1) captures the large-scale, artificially introduced gradient almost entirely. In WFC1, this appears as a smooth left-to-right variation along the x-axis, transitioning from negative weights (purple) on the left to positive weights (yellow) on the right. In WFC2, the gradient is oriented vertically along the y-axis, showing a similar transition from negative to positive weights as one moves from bottom to top. This strong, smooth variation indicates that PCA has isolated the applied gradient as the most significant source of variance across the dataset, consistent with its construction. In contrast, the twentieth principal component (PC20) reveals much finer-scale, higher frequency spatial structure. These patterns lack the smooth gradient seen in PC1 and instead contain more localized vertical striping and pixel-scale variations. This suggests that PC20 and other higher-order components are capturing subtle residual structure, noise, or small systematic differences between frames rather than the dominant gradient. The presence of vertical striping corresponds to readout-related fixed pattern noise or pixel column-dependent bias structures inherent to the detector electronics.

Overall, the separation between PC1 and PC20 demonstrates PCA’s ability to disentangle large-scale systematic trends (the synthetic gradient) from the more complex and fine-scale instrumental signatures. The fact that the gradient is almost perfectly aligned with PC1 for each detector also reinforces the earlier finding that PCA is highly sensitive to global, spatially coherent changes in the superbias frames.

\begin{figure}[H]
	\centering
	\begin{subfigure}{1\textwidth}
		\centering
		\includegraphics[scale=0.075]{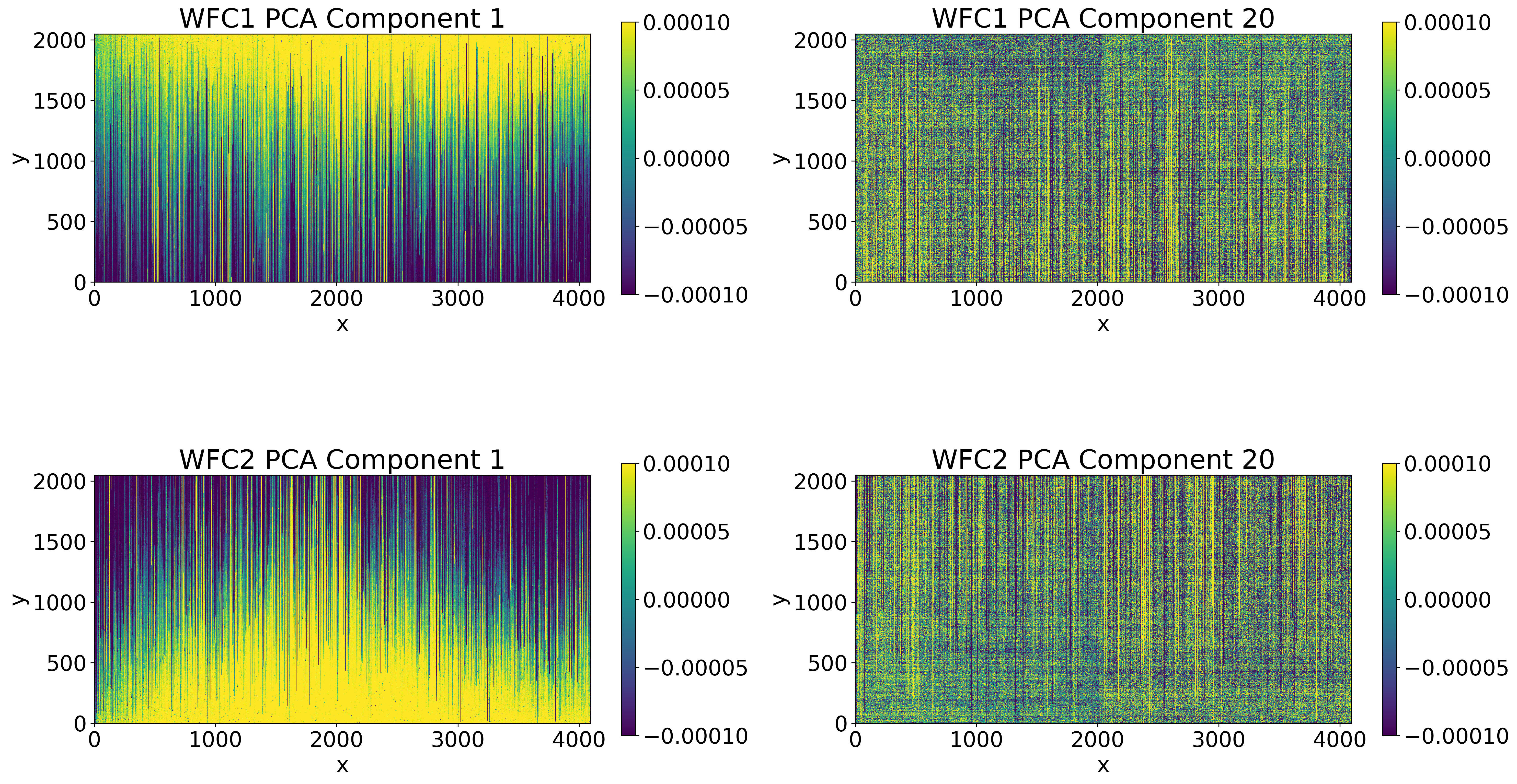}
		\end{subfigure}
\caption{Reconstructed superbias frames from selected principal components for the original 2009 superbias frames. The left column shows the first principal component, and the right column shows the 20th principal component. WFC1 reconstructions are shown in the top row and WFC2 in the bottom row. The readout dark and bias gradient is captured by PC1 while bias striping is captured in later PCs.}
    \label{fig:wfc1_wfc2_components_og_2007}
\end{figure}

We then reconstructed the images of PC1 and PC20 for the original superbias frames. In Figure \ref{fig:wfc1_wfc2_components_og_2007} the structure of PC1 (top-left and bottom-left) appears more complex and less uniform compared to the clean horizontal or vertical gradients seen in the simulated frames. Instead, the primary components here capture variations in bias level across the detector. These structures likely represent hot columns and readout dark in the ACS/WFC. By contrast, WFC1 and WFC2 Component 20 (top-right and bottom-right) are dominated by finer, higher-frequency spatial variations, showing more pixel-scale structure and noise-like patterns. These higher-order components capture less overall variance but reflect localized variations in bias structure which is bias striping in the superbias frames.

Compared to the simulated superbias, where the first principal component was dominated almost entirely by the artificial gradient, PC1 in the original dataset primarily captures detector-specific features such as readout dark patterns and persistent hot columns. The prominence of these structures in PC1 indicates that they represent the most significant and consistent sources of variation across the superbias frames. This observation reinforces the conclusion that, aside from these detector artifacts, the global structure of the superbias files remains remarkably consistent over time. Moreover, finer-scale variations are relegated to higher-order components, suggesting that such subtle features become negligible over time.

\section{Conclusion} \label{s:conclusion}

This study investigated the long-term behavior of ACS/WFC superbias reference files from 2007 to 2024 to assess whether the current strategy of monthly calibration frame delivery remains necessary. Through a combination of statistical trend analysis, synthetic testing, and dimensionality reduction via PCA, we find compelling evidence that the superbias structure in ACS/WFC has remained largely stable since the post-Servicing Mission 4 (SM4) period began in 2009. While the mean bias level has exhibited a slow and steady rise—consistent with readout dark in the WFC detector—no abrupt changes or unexpected trends were detected. Comparisons with WFC3/UVIS show that although ACS/WFC bias levels are higher and evolve more noticeably over time, the changes are linear and predictable.

Mean bias levels increased at rates of $\sim0.063$ $e^-$ per year (WFC1) and $\sim0.051$ $e^-$ (WFC2), faster than the WFC3/UVIS detectors. Our PCA reveals persistent detector-specific patterns, such as readout dark and hot columns, as the dominant sources of variance. The pronounced separation of pre- and post-SM4 data in PCA space underscores the lasting impact of the 2009 electronics upgrade, which cannot be reproduced by gradual trends alone.
Post-SM4 data show time-dependent changes in superbias structure, with signs of stabilization in the most recent years. Importantly, aside from stable fixed-pattern features, large-scale variations remain minimal. While ACS/WFC superbias frames are less stable than WFC3/UVIS, they do not exhibit rapid or unpredictable shifts aside from small-scale features like unstable hot columns. However, given their greater variability due to unstable hot columns and more rapidly increasing readout dark, ACS/WFC calibrations likely still benefit from more frequent superbias deliveries than the annual cadence adopted for WFC3/UVIS.

\section*{Acknowledgements}
We thank Frederick Dauphin and Sean Lockwood for assistance, feedback and support throughout the project. We also thank Jenna Ryon, David Stark and Gagandeep Anand for providing comments to improve the report.
For this report we used the following: \texttt{jupyter} \citep{jupyter}, \texttt{numpy} \citep{cite_numpy}, \texttt{pandas} \citep{pandas}, \texttt{astropy} \citep{astropy}, and \texttt{matplotlib} \citep{matplotlib}.

\bibliography{isr_pca}
\bibliographystyle{apj}

\end{document}